# Characterization of material around the centaur (2060) Chiron from a visible and near-infrared stellar occultation in 2011

A.A. Sickafoose,[1,2,3]* A.S. Bosh,[2,4] J.P. Emery,[5] M.J. Person,[2] C.A. Zuluaga,[2] M. Womack,[6] J.D. Ruprecht,[7] F.B. Bianco,[8,9] and A.M. Zangari[7,10]

[1]*South African Astronomical Observatory, Observatory Rd., Cape Town, 7925, South Africa (Address for correspondence: P.O. Box 9, Observatory, 7935, South Africa)*
[2]*Department of Earth, Atmospheric, and Planetary Sciences, Massachusetts Institute of Technology, 77 Massachusetts Ave., Cambridge, MA 02139, USA*
[3]*Planetary Science Institute, 1700 East Fort Lowell, Tucson, AZ 85719, USA*
[4]*Lowell Observatory, 1400 West Mars Hill Road, Flagstaff, AZ 86001, USA*
[5]*Dept. of Astronomy and Planetary Science, Northern Arizona University, Flagstaff, AZ 86001, USA*
[6]*Florida Space Institute, University of Central Florida, Orlando, FL 32816, USA*
[7]*Lincoln Laboratory, Massachusetts Institute of Technology, Lexington, MA 02421, USA*
[8]*Department of Physics and Astronomy; Joseph R. Biden Jr. School for Public Policy and Administration; Data Science Institute, University of Delaware, Newark, DE 19716, USA*
[9]*Center for Urban Science and Progress, New York University, New York, NY 10003, USA*
[10]*Department of Space Studies, Southwest Research Institute, Boulder, CO 80302, USA*

**ABSTRACT**

The centaur (2060) Chiron exhibits outgassing behaviour and possibly hosts a ring system. On 2011 November 29, Chiron occulted a fairly bright star ($R$~15 mag) as observed from the 3-m NASA Infrared Telescope Facility (IRTF) on Mauna Kea and the 2-m Faulkes Telescope North (FTN) at Haleakala. Data were taken as visible wavelength images and simultaneous, low-resolution, near-infrared (NIR) spectra. Here, we present a detailed examination of the light-curve features in the optical data and an analysis of the NIR spectra. We place a lower limit on the spherical diameter of Chiron's nucleus of 160.2±1.3 km. Sharp, narrow dips were observed between 280–360 km from the centre (depending on event geometry). For a central chord and assumed ring plane, the separated features are 298.5–302 and 308–310.5 km from the nucleus, with normal optical depth ~0.5–0.9, and a gap of 9.1±1.3 km. These features are similar in equivalent depth to Chariklo's inner ring. The absence of absorbing/scattering material near the nucleus suggests that these sharp dips are more likely to be planar rings than a shell of material. The region of relatively-increased transmission is within the 1:2 spin-orbit resonance, consistent with the proposed clearing pattern for a non-axisymmetric nucleus. Characteristics of possible azimuthally incomplete features are presented, which could be transient, as well as a possible shell from ~900–1500 km: future observations are needed for confirmation. There are no significant features in the NIR light curves, nor any correlation between optical features and NIR spectral slope.

## 1 INTRODUCTION

Centaurs are bodies whose orbits have been perturbed by the giant planets – they are considered "planet crossers" and have dynamically short lifetimes (median ~10 Myr, e.g. Tiscareno & Malhotra 2003). These objects are particularly interesting because they are thought to represent a transition between the TransNeptunian Objects (TNOs) and other populations, notably the Jupiter-family comets (JFCs). However, the ties between these populations are not well established. Centaurs are thought to be the primary source for the JFCs (e.g. Levison & Duncan 1997), possibly through a gateway orbit (Sarid *et al.* 2019), and numerical simulations show that they can also evolve outward or be ejected from the Solar System (e.g. Horner *et al.* 2004). Suggested source regions for the centaurs include the Kuiper Belt, Scattered Disk, inner Oort cloud, Plutinos, and Neptune Trojans (Holman & Wisdom 1993, Peixinho *et al.* 2004, Duncan & Levison 1997, Emel'yanenko *et al.* 2005, Horner & Lykawka 2010). In addition, centaur colours have been found to be bimodal, ranging from neutral/blue (like Chiron) to very red (like Pholus) (e.g. Peixinho *et al.* 2003, Tegler *et al.* 2008, Tegler *et al.* 2016). This colour diversity is difficult to match using simple models of surface composition and evolution. A better characterization of individual centaurs can allow more accurate comparison with possible kindred populations as well as an increased understanding of planetary surface processes.

---
* Email: amanda@saao.ac.za

The centaur (2060) Chiron was discovered in 1977 (Kowal 1979). It orbits primarily between Saturn and Uranus, ranging from approximately 8.5 to 19 au, with orbital eccentricity of 0.38 and inclination of 7º. Its rotational period has been reported as 5.92 h (with low error, Marcialis & Buratti 1993), although a slightly shorter rotation of approximately 5.5 h was reported in more recent, sparsely-sampled data (Fornasier *et al.* 2013, Galiazzo *et al.* 2016). Upon discovery, Chiron was thought to be an asteroid, but it has since exhibited outbursting behaviour and is co-designated as comet 95P/Chiron (e.g. Luu & Jewitt 1990, Hartmann *et al.* 1990, Meech & Belton 1990, Jewitt 2009, Bauer *et al.* 2004). Chiron's brightness varies over both short (hour) and long (decade) timescales, and is remarkably uncorrelated with heliocentric distance (e.g. Tholen *et al.* 1988, Belskaya *et al.* 2010).

Successful stellar occultation observations in 1993 and 1994 indicated that Chiron's nucleus is greater than 180 km in diameter (Elliot *et al.* 1995, Bus *et al.* 1996). The Herschel Space Telescope was used to determine Chiron's size as 218±20 km (Fornasier *et al.* 2013), while the most recent measurement is from ALMA (the Atacama Large Milimeter Array) and has a preferred size of 114 × 98 × 62 km (Lellouch *et al.* 2017). The stellar occultations revealed an asymmetric dust coma as well as discrete jet-like features around Chiron's nucleus. Such activity is notable, given the comparatively distant orbit and large nucleus relative to comets. Chiron's visible to near-infrared spectrum is neutral, and water ice has been detected on the surface with temporal variability that is likely due to dilution from a coma (Foster *et al.* 1999, Luu *et al.* 2000, Romon-Martin *et al.* 2003). Polarimetric observations suggest that Chiron's surface is homogeneous, possibly with a small amount of water frost on a dark surface (Belskaya *et al.* 2010). Emission from CN and CO was also detected in amounts consistent with isolated outbursts from a small fraction of Chiron's surface. Such activity could entrain dust or ice particles to create a coma (Bus *et al.* 1991, Womack & Stern 1999, Womack *et al.* 2017).

A stellar occultation by Chiron was observed on 2011 November 29. At the time of this event, Chiron was 17.04 au from Earth and had apparent magnitude 18.8. The occultation data include both visible-wavelength images and low-resolution, near-infrared spectra. Ruprecht *et al.* (2015) reported the first results from this event for the visible-wavelength data. They found symmetric extinction features, 0.7 to 1.0 in optical depth, located roughly 300 km from the centre of Chiron's nucleus. The features were separated by 10–14 km in the sky plane, along the path of the star with respect to Chiron, and were 3±2 and 7±2 km in width, respectively. With sparse occultation chord sampling, it could not be assumed that the features were azimuthally continuous (i.e. a ring system); instead, a near-circular arc or shell of material was proposed.

In 2013, a ring system around the centaur (10199) Chariklo was detected via a stellar occultation observation (Braga-Ribas *et al.* 2014). This system is remarkably similar to the features observed at Chiron: Chariklo is 248±18 km in size (Fornasier *et al.* 2013) and the rings are located 391 and 405 km from the centre of the nucleus, with widths of roughly 7 and 3 km, respectively (Bérard *et al.* 2017, Braga-Ribas *et al.* 2014). Ortiz *et al.* (2015) combined the results from Ruprecht *et al.* (2015) with new rotational light curves and long-term brightness and spectroscopic variations to propose that Chiron has a full ring system similar to Chariklo's. Ortiz *et al.* (2015) also proposed a triaxial shape with a spin-pole orientation that suggest that the variability in the water-ice band may be explained partly by changes in the relative axial tilt of the system.

This manuscript presents the near-infrared data for the 2011 November 29 stellar occultation by Chiron and provides a more detailed analysis of the visible-wavelength data presented in Ruprecht *et al.* (2015), specifically in the context of studying material around the nucleus. The observations are presented in Section 2. Section 3 contains the data analysis and results, including measurement of Chiron's size and a multi-wavelength examination of the occultation light-curve features. The observed flux versus wavelength is derived for the surrounding material, in an attempt to determine particle sizes. Sections 4 and 5 contain a discussion and conclusions.

## 2 OBSERVATIONS

An occultation by Chiron of a $B$ = 15.01 mag, $V$ = 14.45 mag, $R$ = 14.93 mag, $K$ = 13.57 mag star (2UCAC 29938128, colours from the NOMAD catalog Zacharias *et al.* 2004) on 2011 November 29 was predicted to occur. The prediction was based on astrometric data taken over a 3-yr period, utilizing the U.S. Naval Observatory 61-inch and the Lowell 42-inch in Flagstaff, AZ, and the 0.9-m SMARTS (Small and Moderate Aperture Research Telescope System) at Cerro Tololo Inter-American Observatory, Chile. The MIT Ephemeris Correction Model was employed to determine the difference between measured and catalogue positions for the star and Chiron and propagate forward to the occultation time (Elliot *et al.* 2009, Gulbis *et al.* 2010). The final prediction placed the shadow path 147±786 km (3-σ error) north of Mauna Kea, Hawai'i. For that site, the predicted occultation midtime was 08:15:36±00:00:19 UT (1-σ error). The relative velocity was slow for these types of events, at 9.87 km s$^{-1}$.

Observations were made at two sites: NASA's 3-m Infrared Telescope Facility (IRTF) on Mauna Kea, and the 2-m Faulkes Telescope North (FTN) on Haleakala, Maui. The cross-track offset between FTN and the IRTF is 97 km. At the IRTF, data were taken simultaneously as near-infrared (NIR) spectra from SpeX (Rayner *et al.* 2003) and visible-wavelength (< 0.9 μm) images from the MIT Optical Rapid Imaging System (MORIS, Gulbis *et al.* 2011). At the FTN, visible-wavelength images were obtained using the Lucky Imaging and High Speed Photometry (LIHSP) in electron-multiplying (EM) mode. Table 1 contains parameters of the observations, and additional observational details are provided in Ruprecht *et al.* (2015). The occultation datasets were started well in advance of the predicted midtime, with the data analysed here starting at 07:47:55 UT for SpeX, 07:55:00 for MORIS, and 08:03:27

**Table 1.** Dataset parameters for the Chiron occultation on 29 November 2011.

| Telescope and instrument | Observed wavelength (μm) | Cadence (Hz) | Deadtime (s) | Instrument settings | Start and end times (UT) | Signal-to-noise ratio[a] |
|---|---|---|---|---|---|---|
| FTN LIHSP | ~0.6[b] | 4.98 | 0.00674 | clear filter: 10 MHz EM gain 146; 2.4x preamp gain | 08:03:27 08:19:43 | 21 |
| IRTF MORIS | ~0.6[b] | 0.5 | 0.0017 | open filter; 1 MHz conventional; 2.4x preamp gain | 07:55:00 08:24:15[c] | 24 |
| IRTF SpeX | 0.99–2.57[d] | ~0.10[e] | ~4.61[e] | 0.9μm dichroic; 1.6" x 60" slit; 20 SlowCnt; 9 NDR | 07:47:55 08:24:15[c] | 4 |

*Notes:* [a] Over 10 km in the light curve, which is the size of a previously detected Chiron jet feature.
[b] MORIS and LIHSP have the same camera, an Andor iXon, and thus the same CCD response.
[c] Data were taken beyond this point in time, but they were unusable due to clouds.
[d] Over 474 separate channels.
[e] Assuming the mean deadtime for all exposures in the SpeX dataset. The integration time was 5.0 s.

for LIHSP. Conditions were partly cloudy. The star was intermittently visible prior to the occultation datasets, and it disappeared completely by 08:24:15 UT.

## 3 ANALYSIS

### 3.1 Derivation of light curves

Ruprecht *et al.* (2015) contains details of the MORIS and FTN data reductions. Standard calibrations were applied and photometry was performed in order to derive the top two light curves shown in Fig. 1, a magnified version of which is shown in Fig. 2 for the FTN data.

For the IRTF SpeX data, signals from the occulted star and a comparison star were obtained by making and applying a flatfield and a bad-pixel mask, applying a linearity correction, straightening the spectral smile, and subtracting background flux. Spectra for each exposure were then extracted in terms of counts per column. The occultation star signal was normalized by dividing by the comparison star signal on each frame. The SpeX data have 474 distinct wavelength channels. To exclude noisy channels, and allow the widest wavelength coverage possible, only channels 10 through 453 (1.02–2.50 μm) are used in these analyses. The median SpeX light curve over the usable wavelength channels is shown at the bottom of Fig. 1.

### 3.2 Model fits to Chiron's nucleus

To determine characteristics from the occultation by Chiron's nucleus observed at the IRTF, we use a square-well model for a stellar occultation by a body without an atmosphere. The model follows that described in Elliot *et al.* (2010), which includes the finite integration time of the instrument and considers cases where the integration bins straddle immersion and emersion. For long cycle times, data points at fractional flux levels can cause the model fit to appear misleading because they do not accurately represent the occultation. Thus, it is useful to consider the derived square-well occultation component. To illustrate, Fig. 3 shows the square-well fit to the median SpeX light-curve data. The results of the model fits to the SpeX data and the higher-time-resolution MORIS data are listed in Table 2. It is instructive to know the accuracy levels for the derived square-well-model parameters, given the different observing cadences and data quality for the two instruments that simultaneously obtained data at this site.

### 3.3 Event geometry

Only one location detected an occultation by Chiron's nucleus, and poor weather prevented data collection before and after the event when Chiron and the star were well separated. Therefore, the geometry of the event is not accurately established. As presented in Ruprecht *et al.* (2015), Chiron's radius is between 72 and 97 km if the IRTF chord were central. If the chord were north of the centre, then the data can be consistent with the maximum proposed size for Chiron of 156 km in radius, which is the 3-$\sigma$ upper limit from Elliot *et al.* (1995). These cases set the limiting minimum and maximum possible nucleus sizes. In Fig. 4, schematic diagrams of Chiron are plotted along with the locations of the light-curve data for these two geometries. The IRTF chord cannot be south of centre, since there was no nucleus occultation detected from the FTN.

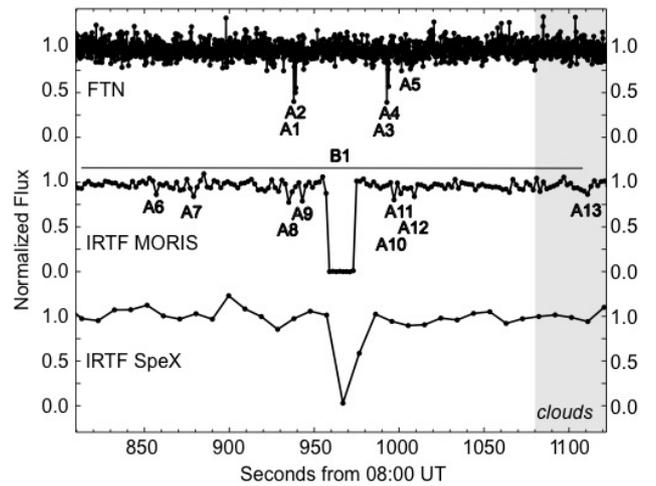

**Figure 1.** Light curves for the 2011 November 29 occultation by Chiron. Data points are at the midtime of each exposure. The SpeX data are the median values over the useful wavelength range 1.02–2.50 μm and show no obvious features other than the occultation by the nucleus. The FTN detected steep dips before and after the midtime, but no solid-body occultation. MORIS detected an obvious occultation by the nucleus, as well as small dips before and after. Significant, narrow features are labelled A1–A13 (A5 is only significant when the data are binned, see Fig. 5). A broad, low-level feature, the extent of which is shown by a line labelled B1, is discussed in Section 4 and shown in Fig. 9. Data in and beyond the grey region show degradation due to clouds (see Sections 3.4 and 3.5).

**Table 2.** Results from square-well fits to the occultation by Chiron's nucleus.

| Instrument | Chord duration (s) | Chord length (km) | Occultation midtime (UT) |
|---|---|---|---|
| SpeX | 13.34±3.26 | 131.7±32.2 | 08:16:08.95 ± 00:00:03.65 |
| MORIS | 16.23±0.13 | 160.2±1.3 | 08:16:05.86 ± 00:00:00.14 |

Note that occultation observations directly measure distances in the sky plane. If the pole orientation of the nucleus is well established, then those measurements can be converted to radial distance from the centre of the body. Unless otherwise noted, all spatial values presented here are in the sky plane. Notable exceptions are in some columns of Table 3 and in Section 4, where we assume the preferred ring-pole solution for Chiron proposed by Ortiz *et al.* (2015) and use distances in the ring plane, to allow comparison of significant light-curve features with the Chariklo system.

### 3.4 Significant features in the light curves

To identify the locations of features in the light curves, the variation from the baseline is examined. Well outside of the occultation, the standard deviation is 0.076 normalized flux for the SpeX data, 0.029 for the MORIS data, 0.087 for the full-resolution FTN data, and 0.033 for a ten-point running average of the FTN data. The running average of the FTN data is considered because it matches the time resolution of the MORIS data. Four-sigma variations in the light curves are indicated in Fig. 5, for each of the considered geometries: other than the occultation by the nucleus, there are no significant features identified in the NIR SpeX light curve, so it is not displayed in this plot. The features are tested for symmetry by overlaying the immersion and emersion portions of the light curves (which can reveal continuation on either side of the nucleus), and by plotting the MORIS and FTN data together in terms of distance from the centre (which can reveal azimuthal continuity around the nucleus). For better visualization, these 4-$\sigma$ locations are represented in Fig. 4 by the large dots in the sky-plane geometry.

To characterize the features, we employ a square-well diffraction model for occultation profiles that treats the blocking material as a uniformly transmitting grey screen with abrupt edges (Elliot *et al.* 1984). The important size scales for diffraction effects are the Fresnel length ($\sqrt{\lambda D/2}$, where $\lambda$ is the wavelength of the observation and $D$ is the distance from Earth), the stellar diameter, and the integration length of the observations (integration time multiplied by the velocity in the sky plane). Here, the Fresnel length is 0.62 km, the maximum stellar diameter is estimated to be <0.1 km (using NOMAD magnitudes and the equations for a giant star from van Belle, 1999), and the lengths of the integrations are 1.97 (FTN), 19.74 (MORIS), and 98.7 km (SpeX). Thus, diffraction effects for these observations are dominated by the integration times and we neglect the stellar profile.

For each of the labelled narrow features A1–A13, characteristics are provided in Table 3. Sky-plane distances and line-of-sight optical depths ($\tau$) are given to provide direct measurements. To allow comparison with other ring systems, corresponding ring-plane values, normal optical depths ($\tau_N$), and equivalent widths are given, assuming a central chord and the opening angle for the preferred solution of a proposed Chiron ring system from Ortiz *et al.* (2015).

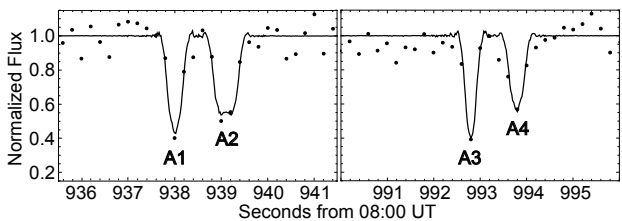

**Figure 2.** Expanded view of the FTN light curve to more clearly show the distinct dips on either side of the predicted midtime. Dots represent the data and solid lines are diffraction-model fits to the data with parameters given in Table 3. The significant features are labelled A1–A4, corresponding to the same labels in Figs. 1, 4, and 5.

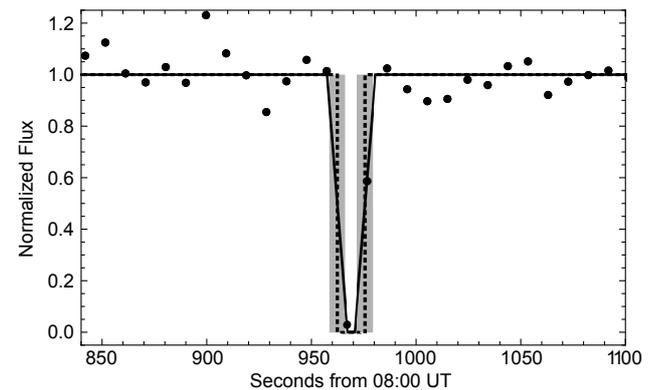

**Figure 3.** Square-well model fit to the SpeX data. Dots represent the data and the solid line is the fit, allowing for data points that straddle immersion and emersion. The dashed line is the corresponding square-well occultation component of the model, which is what would be observed without averaging over the instrument integration time. The grey regions represent the errors of the fit for immersion and emersion times, as listed in Table 2.

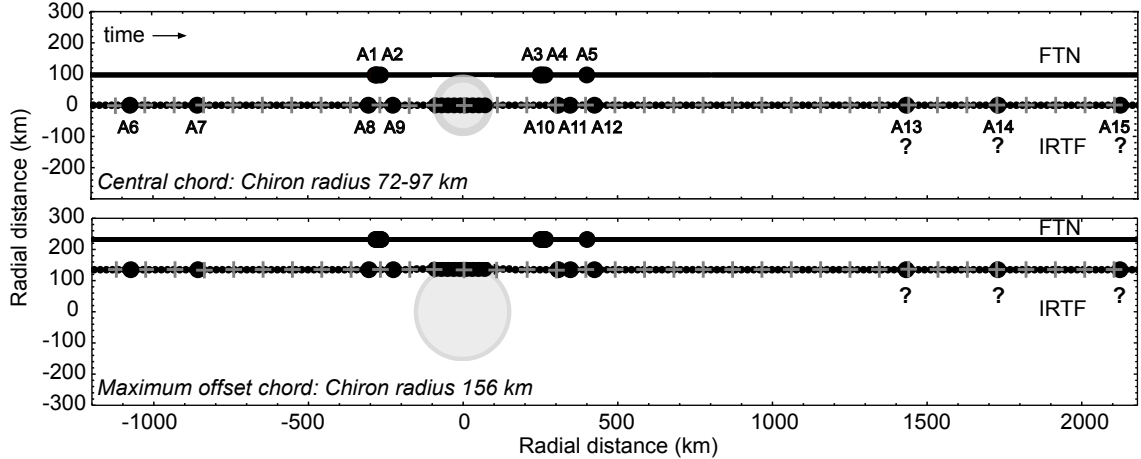

**Figure 4.** Sky-plane geometries of Chiron with the occultation data for limiting cases: (*top*) the IRTF as the central chord and (*bottom*) the maximum possible chord offset from centre. The darker grey shading around the nucleus represents (*top*) the range of sizes from Ruprecht *et al.* (2015) and (*bottom*) the maximum size from Elliot *et al.* (1995) with the error on the IRTF chord length as the lower boundary. Each small black dot represents the midpoint of an image taken from MORIS on the IRTF and the LIHSP at the FTN (the FTN images are so closely spaced that they appear as a solid line). The grey crosses represent midpoints of low-resolution spectra taken with SpeX on the IRTF. Large black dots indicate places in the occultation light curves where the data are more than four sigma below the baseline. Note that all of the data during the occultation by the nucleus meet this criterion. Significant, narrow features are labelled A1-A15 on the top plot and are also indicated in Figs. 1, 2, and 5. The question marks indicate features that are likely compromised by cloud. For simplicity, the nucleus is shown here to be spherical, although Ortiz *et al.* (2015) and Lellouch *et al.* (2017) proposed it is ellipsoidal.

In Fig. 5, A8 and A10 are 304–336 km from the centre in the MORIS data and A1+A2 and A3+A4 are 274–360 km in the averaged FTN data. These features are detected at similar locations on either side of the nucleus as well as azimuthally around the nucleus, in the regions probed by the chords. In the FTN full-resolution data shown in Fig. 2, these features separate into dips at 291–361 km (A1 and A4) and 280–353 km (A2 and A3). These dips were described in Ruprecht *et al.* (2015) and were the basis of the proposed Chiron dual-ring system in Ortiz *et al.* (2015). In the full resolution data, the widths of A1 and A4 are smaller on immersion than emersion (3 to 3.6 km), the opposite of the A2 and A3 (4.7 to 2.8 km). The optical depths vary between $\tau \sim 0.6$ and 1 ($\tau_N \sim 0.5$–0.9), such that the equivalent width stays nearly constant at 1.4–1.5 km for A1 and A4 and 1.5–2.0 km for A2 and A3. The MORIS and averaged FTN data have a lower sampling cadence and encompass both of the dips in one feature, with $\tau \sim 0.15$–0.31.

An additional feature is detected by both MORIS and FTN, at 425–447 (A12) and 420–471 km (A5), respectively, but it is only observed during emersion. The lack of appearance on immersion suggests an extended arc that does not fully encircle the nucleus. A12 has optical depth $\tau \sim 0.2$ and A5 has lower optical depth ($\tau \sim 0.07$), and they are not apparent in the full resolution FTN data. The equivalent width is larger in the MORIS data than in the FTN data, implying an azimuthal density variation even along the detected arc, although the errors are large.

Seven features are identified in the MORIS data that have no match between immersion and emersion nor in the FTN data. Two features are located near the distinctive FTN dips: slightly interior on immersion (225–263 km, A9) and slightly exterior on emersion (346–373 km, A11). Two features occur on immersion at a moderate distance from the nucleus, 1074–1083 km (A6) and 857–868 km (A7). The final three, A13–A15, occur on emersion at distances greater than 1400 km from the centre.

The IRTF data were stopped at ~1455 s from 08:00 UT because of signal loss due to clouds. The lower signal-to-noise ratio (SNR) of the SpeX data rendered the target undetectable by that instrument at ~1440 s. After ~1187 s, the reference star wandered off the frame on the FTN frames. Although there is no obvious effect in the normalized light curves in Fig. 1, the raw signal began degrading at ~1080 s. The noise increased in the MORIS and SpeX light curves from 1080 to 1440 s, with a standard deviation at these times of 0.05 and 0.15, respectively. Also, the NIR data show a reddening in spectral slope during that time (see Section 3.5). Therefore, we suspect that data after ~1080 s is contaminated by clouds and have labelled these distant emersion features with question marks in Figs. 4 & 5.

Finally, there is another feature in Fig. 5 that stands out in the averaged FTN data, interior to the steep dips, at ~204.2–205.9 km from the centre assuming a central IRTF chord and ~293.4–294.7 km from the centre for an offset chord (errors on all values of ±9.9 km). It is not labelled nor listed in Table 3 because it does not meet the 4-$\sigma$ requirement. This feature is more significant on emersion (3$\sigma$) than immersion (2.2$\sigma$). On emersion, the width is 28.9±0.4 km, $\tau$=0.11±0.04 ($\tau_N$=0.09±0.03), and the equivalent width is 2.3±0.4 km. The significance level of this feature is lower than the others discussed here, but it is worth noting. This could be another azimuthally incomplete arc, possibly connected with the MORIS feature closest to the nucleus that was detected on immersion with equivalent width 4.4±0.9 km.

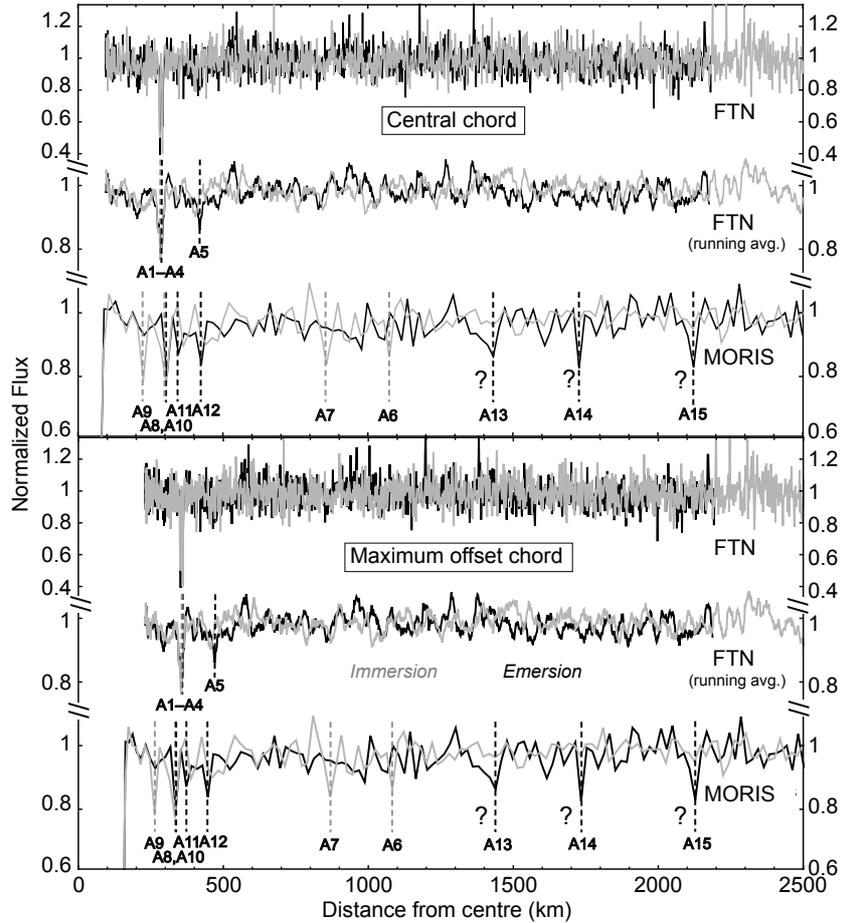

**Figure 5.** Investigation of significance and symmetry in light curve features. The distances in this plot assume the two limiting geometries shown in Fig. 4, with the top panel being the minimum possible Chiron nucleus size and the bottom panel being the maximum nucleus size. The location of the chords is likely somewhere between these two geometries. FTN data are shown at full time resolution as well as a running average over 10 points to match the time and distance resolution of MORIS. For each light curve, immersion (grey) and emersion (black) data are overlaid. The dashed lines indicate data points that are more than four sigma from the baselines: indicator lines are not used for the full-resolution FTN data because the features are obvious. Significant, narrow features are labelled A1-A15 on the top plot and are also indicated in Figs. 1, 2, and 4. Question marks indicate features that are likely compromised due to clouds.

### 3.5 SpeX flux versus wavelength

The SpeX data contain a wealth of information in wavelength space, although they are at fairly low SNR. We remove data outside of the optimal NIR sky passbands, keeping that in the ranges of the standard $J$ (1.164–1.326 μm), $H$ (1.483–1.779 μm), and $K$ (2.027–2.363 μm) filters. To prevent skewing the results, data points with anomalously high or low errors are also removed (normalized flux error > 0.2 or < 0.015; a total of 6 per cent of the sample).

Particles scatter and absorb light as a function of observed wavelength; therefore, attenuation of stellar flux by material around Chiron could manifest in the SpeX dataset as a trend of flux versus wavelength (see, for example, Gulbis et al. 2015). Light scattering is most sensitive to particles on the order of the observed wavelength, so these observations probe for micron-sized particles, which have been previously observed in Chiron's coma (Luu & Jewitt 1990). Generally speaking, for water-ice particles over the SpeX wavelength range, positive spectral slopes would be caused by submicron-sized particles and negative spectral slopes would be caused by larger particles. To check for such trends, a linear function is fit to the SpeX data of flux versus wavelength. The reference wavelength for the fits is 1.63 μm, the middle of $J$ band, and the fits are weighted by the error on each data point. Tests were carried out for binned data in each wavelength and over time. The optimal choice is to analyse each time step (to maintain the highest possible spatial resolution) and bin by three in wavelength space to increase SNR while maintaining spectral resolution.

In Fig. 6, the slopes from linear fits to the NIR flux versus wavelength are plotted at each time step, along with these and MORIS light curves. The mean slope value is fairly constant at –0.05±0.06 flux μm$^{-1}$ until ~1080 s, after which the mean slope settles at a steeper value of –0.12±0.08 flux μm$^{-1}$. Clouds would cause the spectral slope to become more negative because of the water bands

**Table 3.** Characterization of 4-σ features in the optical light curves.

| Feature label[a] | Midtime[a] (s from 08:00 UT) | Distance from centre[b] (km; central chord) | Distance from centre[b] (km; offset chord) | Sky-plane width[c] (km) | Line-of-sight optical depth[c], $\tau$ | Ring-plane distance from centre[d] (km; central chord) | Ring-plane width[d] (km) | Normal optical depth[e], $\tau_N$ | Equivalent width[f] (km) |
|---|---|---|---|---|---|---|---|---|---|
| *MORIS data* | | | | | | | | | |
| A6 | 857.0 | 1074.5 | 1083.0 | 16.2±9.8 | 0.16±0.06 | 1078.4 | 16.3±9.8 | 0.14±0.05 | 2.0±1.4 |
| A7 | 879.0 | 857.4 | 867.9 | 33.2±3.0 | 0.18±0.05 | 860.5 | 33.3±3.0 | 0.15±0.04 | 4.8±1.0 |
| A8 | 935.0 | 304.6 | 333.2 | 29.4±1.8 | 0.26±0.05 | 305.7 | 29.5±1.8 | 0.22±0.04 | 5.9±0.8 |
| A9 | 943.0 | 225.7 | 263.0 | 17.5±1.7 | 0.34±0.09 | 226.5 | 17.6±1.7 | 0.29±0.08 | 4.4±0.9 |
| A10 | 997.0 | 307.3 | 335.6 | 11.1±4.8 | 0.31±0.04 | 308.4 | 11.1±4.8 | 0.27±0.03 | 2.6±1.2 |
| A11 | 1001.0 | 346.8 | 372.1 | 19.8±1.6 | 0.18±0.06 | 348.1 | 19.9±1.6 | 0.15±0.05 | 2.8±0.6 |
| A12 | 1009.0 | 425.7 | 446.6 | 11.6±7.1 | 0.20±0.04 | 427.2 | 11.6±7.1 | 0.17±0.03 | 1.8±1.2 |
| A13 | 1111.0[g] | 1432.5 | 1438.8 | 37.6±2.2 | 0.16±0.04 | 1437.7 | 37.7±2.2 | 0.14±0.03 | 4.8±0.8 |
| A14 | 1141.0[g] | 1728.6 | 1733.9 | 11.1±3.8 | 0.24±0.04 | 1734.8 | 11.1±3.8 | 0.21±0.03 | 2.0±0.5 |
| A15 | 1181.0[g] | 2123.4 | 2127.7 | 29.3±2.4 | 0.19±0.04 | 2131.0 | 29.4±2.4 | 0.16±0.03 | 4.4±0.8 |
| *FTN data (10-point running average)* | | | | | | | | | |
| A1+A2 | 937.2–939.0 | 274.0–290.8 | 345.7–359.1 | 27.7±2.0 | 0.15±0.05 | 292.5–308.2 | 25.8±1.9 | 0.13±0.04 | 3.1±0.7 |
| A3+A4 | 991.2–992.8 | 276.5–291.4 | 347.6–359.6 | 26.9±4.6 | 0.15±0.05 | 294.8–308.7 | 25.1±4.2 | 0.13±0.04 | 3.0±0.9 |
| A5 | 1006.5 | 420.8 | 470.6 | 20.4±7.4 | 0.07±0.05 | 432.32 | 19.8±7.2 | 0.06±0.04 | 1.1±0.7 |
| *FTN Data (full resolution)* | | | | | | | | | |
| A1[h] | 938.0 | 291.7 | 359.9 | 3.0±0.3 | 0.87±0.10 | 309.0 | 2.8±0.3 | 0.75±0.09 | 1.5±0.2 |
| A2[h] | 939.0–939.2 | 280.5–282.4 | 350.9–352.3 | 4.7±0.6 | 0.70±0.18 | 298.6–300.3 | 4.4±0.6 | 0.60±0.15 | 2.0±0.4 |
| A3[h] | 992.8 | 283.0 | 352.8 | 2.8±0.4 | 0.99±0.29 | 300.9 | 2.6±0.4 | 0.85±0.25 | 1.5±0.3 |
| A4[h] | 993.8 | 292.3 | 360.3 | 3.6±0.6 | 0.60±0.20 | 309.6 | 3.4±0.6 | 0.51±0.17 | 1.4±0.4 |

*Notes:* [a]Corresponding to the colour scheme in Fig. 5, rows with grey midtimes are immersion features, black are emersion.
[b]Distance from the centre of the nucleus, as measured in the sky plane. Error is ± half of a time step: 1 s or 9.9 km for MORIS and averaged FTN data; 0.1 s or 0.9 km for FTN full-resolution data.
[c]Derived from least-squares fits to the data using a square-well diffraction model for occultation profiles (Elliot *et al.* 1984), as described in Section 3.4.
[d]Sky-plane locations are converted to the ring plane assuming a central chord and the preferred pole position from Ortiz *et al.* (2015).
[e]Normal optical depths are determined assuming an opening angle of $B=59°$, where $\tau_N = \tau \sin|B|$ for a polylayer ring. This angle is from the preferred ring pole solution of Ortiz *et al.* (2015).
[f]Equivalent width is the width in the ring plane multiplied by opacity, where opacity = $1-e^{-\tau_N}$.
[g]Features that may be compromised due to cloud, which are indicated by question marks in Figs. 4 and 5.
[h]Features that were reported in Ruprecht *et al.* (2015).

near 2 μm: this change in slope, combined with the increased noise in raw signal and the light curves as discussed in Section 3.3, suggests that the data after ~1080 s are contaminated by clouds. Data from the CFHT SkyProbe confirm a sharp increase in attenuation (~0.4 mag) at this time, as well as further 1-2 magnitude attenuations after we lost signal at 08:24 UT.

The mean slopes before and after 1080 s are considered the baseline values, containing flux from the star and Chiron. Chiron's spectral slope, that of the data point during the occultation, is found to be 0.02±0.02 flux μm$^{-1}$ (Fig. 7). It has a more positive slope than the surrounding baseline, although both are consistent with zero. This neutral slope matches that found for previous NIR observations of Chiron (e.g. Romon-Martin *et al.* 2003).

Extinction of light by particles should cause a decrease in flux in both visible (MORIS) and NIR (SpeX) wavelengths. Thus, if there is scattering, correlated dips in the two light curves should be observed. Locations where the SpeX exposures overlap 4-σ features in the MORIS light curve are highlighted in Fig. 6 as open circles. For each of these locations (before the clouds), the flux versus wavelength data and weighted linear fits are provided in Fig. 7. These plots also serve to demonstrate how the spectral slopes were derived at each time step to create Fig. 6.

None of the SpeX data points aligned with features in the optical light curve have significant spectral slopes; potential variations are at only 1–2σ. We conclude that any trends in flux versus wavelength in the dataset are within the noise and the data cannot be used to investigate particle sizes of occulting material.

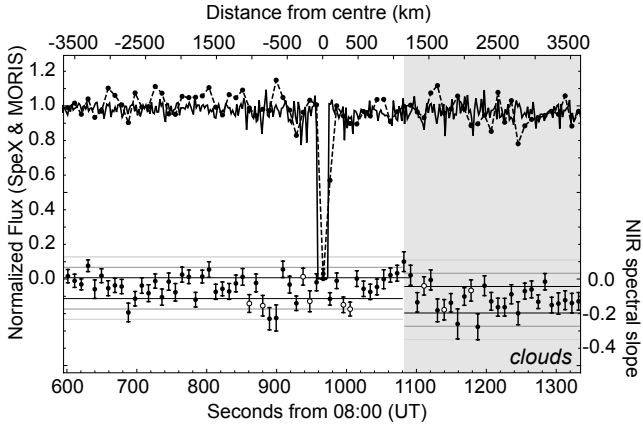

**Figure 6.** Slopes from linear fits to SpeX NIR flux versus wavelength data, along with the SpeX and MORIS light curves. For reference, the straight lines represent ±1, 2, and 3 σ (darker to lighter) from the mean of the NIR flux slope well before and after the occultation. The change in NIR flux slope around ~1080 s is apparent, and the data before and after this point are separately used to determine the sigma levels. The SpeX and MORIS light curves are overplotted to allow direct comparison of features: SpeX data are the dashed line with points and MORIS data are the solid line. Open points in the spectral slope data represent locations where SpeX exposures overlap the significant MORIS features identified in Figs. 4 & 5 and Table 3.

## 4 DISCUSSION

An occultation by Chiron's nucleus was observed simultaneously from two instruments, MORIS and SpeX, at different cadences. The cadence of SpeX was a factor of 5 slower than MORIS, with significantly larger deadtime. For each dataset, we employ a square-well fit to determine the chord duration and occultation midtime. We find durations of 160.2±1.3 km for the MORIS data and 131.7±32.2 km for the SpeX data. Ruprecht *et al.* (2015) found a chord length of 158±14 km for the MORIS data, based on the errors of the individual data points at immersion and emersion. The results for chord lengths and midtimes are consistent within the error bars. The efficacy of the square-well model is validated for low-cadence occultation data, and the most accurate result is obtained by applying the model to the data with the highest time resolution. This example is useful for future occultation observation planning, to estimate the effect of observing cadence on the accuracy of derived chord lengths and midtimes.

Detection of the nucleus from only one site prevents accurate determination of the event geometry. We consider two limiting cases of the IRTF chord being central (minimum nucleus size) or on the northern edge (maximum nucleus size). The true geometry likely falls between these cases. Therefore, the measured chord length places only a lower limit on Chiron's size. This size measurement is consistent with previous occultation data and thermal measurements for the diameter of a spherical Chiron nucleus, which range from 142±10 km to 218±20 km (Fornasier *et al.* 2013, Fernandez *et al.* 2002, Groussin *et al.* 2004, Lebofsky *et al.* 1984, Altenhoff & Stumpff 1995, Campins *et al.* 1994, Lellouch *et al.* 2017). However, the lower limit from the occultation (160.2 km) is substantially larger than the largest axis (114 km) for the favoured, elliptical model from Lellouch *et al.* (2017) but is consistent with the spherical models.

A number of significant, >4σ, features are identified in the visible-wavelength light curves. The following cases are evaluated when analysing the symmetry of the light-curve features:

(i) at similar locations on immersion and emersion, and in both datasets (as for a consistent ring system encircling the nucleus);

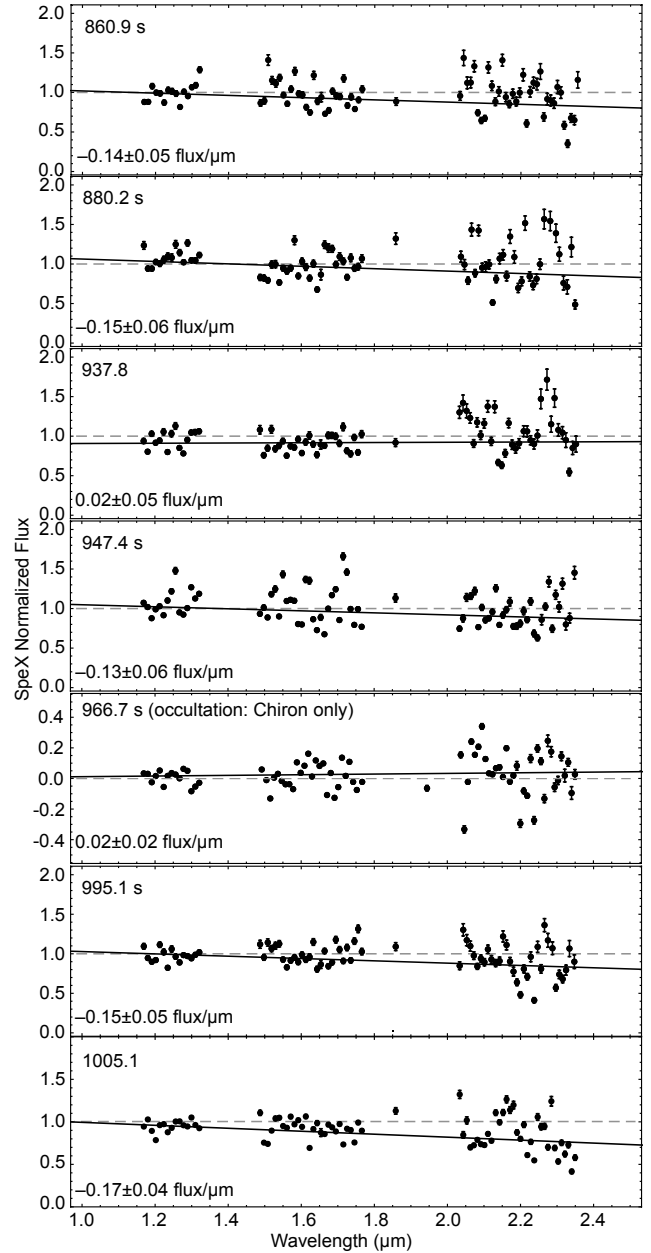

**Figure 7.** Flux versus wavelength with linear fits for SpeX data points that overlap significant MORIS light curve features, prior to 1080 s. The single SpeX time step during the occultation, containing flux from Chiron only, is also included. The wavelength data, binned by three, are plotted with the least-squares linear fits (solid lines). A grey, dashed line is provided at full flux with zero slope, for comparison. The times and fitted spectral slopes are listed in the plots, with the spectral slopes (excluding Chiron only) plotted as open circles in Fig. 6.

(ii) at similar locations on immersion and emersion, but not in both data sets (as for a ring that has clumps or is otherwise azimuthally incomplete);
(iii) at similar locations between datasets but not on immersion and emersion (as for an arc of material that does not encircle the whole nucleus), and
(iv) one-off features that do not align during immersion and emersion or with the other dataset (as for an isolated clump of material).

We find one feature that matches (i): A8 and A10 in the MORIS data, corresponding to A1+A2, A3+A4 in the FTN averaged data. There is one feature that matches (iii): A12 in the MORIS data, corresponding to A5 in the FTN averaged data. There are four or more features that match (iv): A6, A7, A9, A11, possibly A13–A15. A visual representation of the identified features, with distances in Chiron's proposed ring plane, is provided in Fig. 8.

A detailed characterization is provided of A1–A4, A8, and A10, which were described in Ruprecht *et al.* (2015) and proposed to be rings by Ortiz *et al.* (2015) (with A1 and A4 being an outer ring, and A2 and A3 being an inner ring). These features are located between 282–352 km and 292–360 km from the centre in the sky plane, with exact value depending on the event geometry. If we assume the ring-pole solution from Ortiz *et al.* (2015), and a central chord, the outer ring (A1 & A4) is at 298.5–302 km and the inner ring (A2 and A3) at 308–310.5 km. The gap between the centres of the rings is 9.1± 1.3 km. The amount of material in both rings is consistent azimuthally, with equivalent widths agreeing in the error bars at ~1.6 km; however, the width of each feature changes from ~3 km to ~4–5 km (essentially, the material clumps and spreads). The results for the widths of these proposed rings, and the gap between them, differ slightly from those reported by Ruprecht *et al.* (2015), in which the measurements were based on looking at individual data points rather than diffraction-model fits.

As shown in Fig. 8, these proposed rings are broadly similar to Chariklo's two-ring system in location, extent, and gap size (Braga-Ribas *et al.* 2014). However, if they fully encompass Chiron, these features are closer to the nucleus by 50–100 km. Unlike Chariklo's relatively azimuthally-congruent, more substantial inner ring ($\tau_N$~ 0.4) and more optically thin outer ring ($\tau_N$~ 0.06), the proposed rings at Chiron are like each other. Chiron's A1–A4 match Chariklo's inner ring in equivalent depth and the A1 & A4 and A2 & A3 radial widths vary by similar amounts at different azimuths, similar to the variations reported for Chariklo's inner ring (Bérard *et al.* 2017). The detection limits (>4σ) on this dataset are $\tau_N$~ 0.05 for 20-km features and $\tau_N$~ 0.18 for 2-km features; therefore, a feature like Chariklo's outer ring would not have been detected.

The new features described here include an arc at ~430 km with equivalent width varying between the two sites from ~1.1 to 1.8 km (A5 and A12) and one-off material at ~230, 350, 860, and 1075 km with equivalent widths from 2.0–4.8 km (A9, A11, A7, and A6). The arc is exterior to the proposed rings but similar to the combined detections of these features (A8 and A10) in terms of size, optical depth, and azimuthal variation. The four, one-off features are thinner and have lower optical depths than those previously detected, which could be the result of the higher spatial resolution of this dataset. As concluded for previous Chiron occultations, the most likely explanation for these features is dust entrained in narrow jets or concentrated outgassing (Bus *et al.* 1996, Elliot *et al.* 1995). Observations of Chiron in 2011 December showed a light-curve amplitude of 0.06–0.07, suggesting moderate cometary activity near the time of the occultation (Fornasier *et al.* 2013).

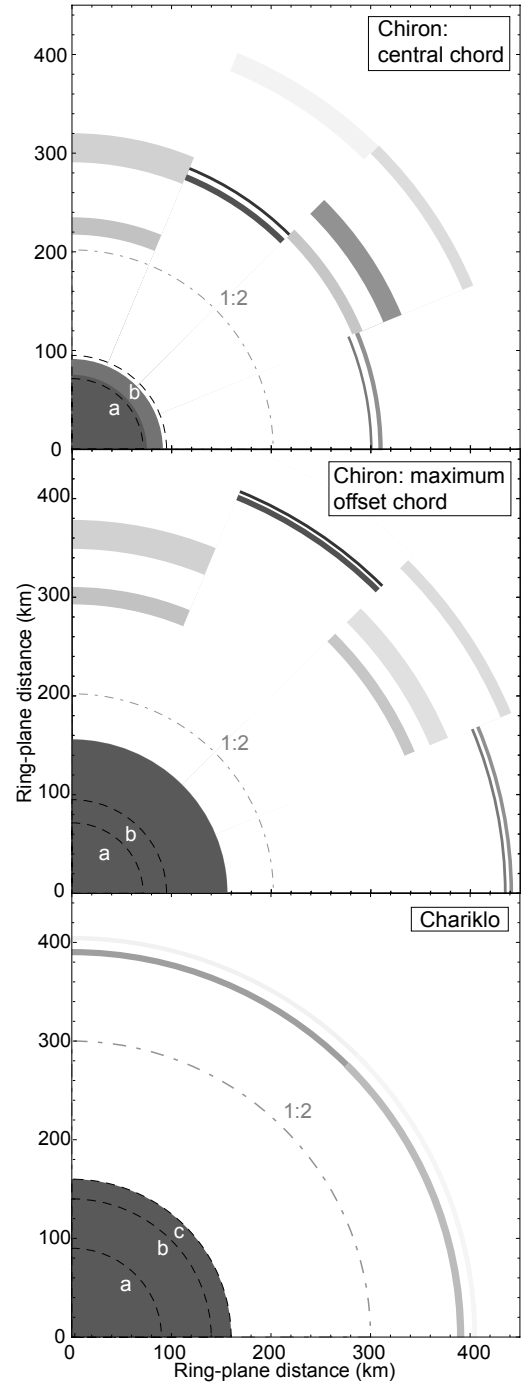

**Figure 8.** Schematic diagrams of the Chiron and Chariklo systems. The features detected in this work are shown in the top two panels, which correspond to the two limiting geometries for the occultation as shown in Fig. 4. Nucleus sizes are plotted in dark grey. Dotted lines indicate proposed ellipsoidal semi-axes $a$ and $b$ from Ortiz *et al.* (2015) for Chiron and $a$, $b$, and $c$ from Leiva *et al.* (2017) for Chariklo. Features are plotted at the locations and with widths given in Table 3, including error bars on widths, and the greyscale corresponds to optical depth. Moving clockwise from the top left in each panel, the data are immersion to emersion, with MORIS followed by FTN for Chiron. The distance scale is in the ring plane. For Chiron, the preferred pole orientation from Ortiz *et al.* (2015) is used. For reference, the location of the 1:2 resonance is marked as a grey dot-dashed line and calculated as $2^{2/3}(GM/\Omega^2)^{1/3}$, where $G$ is the gravitational constant, $M$ is the mass of the nucleus, and $\Omega$ is the spin rate (5.918 hr for Chiron and 7.004 hr for Chariklo). Note that for an ellipsoidal body, this would be the 2:4 resonance (Sicardy *et al.* 2019).

The stellar occultations detected by Chiron in 1993 and 1994 do not show azimuthally continuous features like those observed in 2011. As noted by Ortiz *et al.* (2015), the lack of accurate geometric solutions for these events allows room for speculation. For the 1993 event (Bus *et al.* 1996), a single feature, A4 from Site 4, is closest to the expected location, size, and optical depth of the proposed rings; however, no corresponding feature was detected in that chord on emersion. Site 4 had integration distances of 18 km per data point, which is similar to the MORIS dataset in 2011, while the other sites were lower resolution and would not necessarily have been able to detect the proposed rings. From the 1994 event (Elliot *et al.* 1995), a single feature F1 could have been a nucleus graze or a ring detection (at sky-plane width 4.5–9.2 km and $\tau$=0.92–0.96). The resolution of the two chords that detected this feature are similar to that of the FTN data in 2011, at 7–10 km. Again, the geometry is difficult to reconcile. If F1 were a nucleus graze, the rings should have been detected on both immersion and emersion in both chords: instead, one broader, fainter feature (F2 at 74 km at $\tau$=0.11) was detected in emersion only. If F1 were a ring graze, it is more likely to be slightly extended than such a sharp dip. Any grazing geometry is statistically not very likely to occur. Our conclusion is that the proposed ring system was not in the same state in the 90s as present day.

Multiple investigations have determined that Chiron has transient outbursts, with different characteristics and including development of a coma (Luu 1993, Belskaya *et al.* 2010, Ortiz *et al.* 2015, Womack *et al.* 2017). Re-analyses of optical data of Chiron's dust coma from 2007-2008 (Tozzi *et al.* 2012) and Herschel PACS 70 and 100 μm data are consistent with a dust production rate of at least 10 and up to 45 kg/s (Fornasier *et al.* 2013). Fornasier *et al.* (2013) propose that the observations can be explained by dust grains of size < 0.14 μm being lifted from the surface by significant volatile production, ranging from $Q$(gas)~(1–10)$\times 10^{27}$ molecules/sec, with $CO_2$ and CO nominated as viable candidates. While no detections, or significant upper limits, exist for $CO_2$, we point out that CO emission was detected in Chiron at 8.5 au near perihelion at roughly this production rate (Womack & Stern 1999).

It is unclear how such outgassing occurs. One proposed mechanism of Chiron's activity is the release of highly volatile species, such as CO, $CO_2$, $N_2$ or $CH_4$, which are trapped in an amorphous water-ice matrix that undergoes crystallization (Sarid *et al.* 2005). This process would not strictly rely on insolation in a continuous way, but could occur sporadically, as different pockets of amorphous water ice are heated to sufficient amount to trigger the process. Such a response would be largely unpredictable and is unlikely to follow an inverse-square-law dependence on heliocentric distance, which is in agreement with what is observed for Chiron's activity. For most of its orbit, Chiron is far enough from the Sun that its equilibrium blackbody temperature is considered to be outside the temperature range where amorphous water ice can readily undergo this phase change; however, thermal infrared observations indicate that Chiron is warmer than predicted by simple thermal equilibrium models, at T=125–140K (Campins *et al.* 1994). In principle, this temperature is high enough to trigger the phase change in amorphous water ice. Alternately, Chiron's typically-observed dust-production rate is partly suggested to be caused by material from Chiron's proposed ring system, which falls and impacts the surface, triggering ejection from the surface (Ortiz *et al.* 2015, Cikota *et al.* 2018). Given Chiron's fast rotation and possible low density, rotational disruption is another process that, while unlikely, should be explored. This process could loft material, including small boulders, into orbit, which could then be shredded if within the Roche limit (Pan & Wu 2016, Kokotanekova *et al.* 2017). Accurate determination of Chiron's density, composition, and structure would help ascertain how likely a contributor rotational disruption is to a coma (Toth & Lisse 2006). It is also possible for more than one process to be playing an important role in Chiron's intermittent activity.

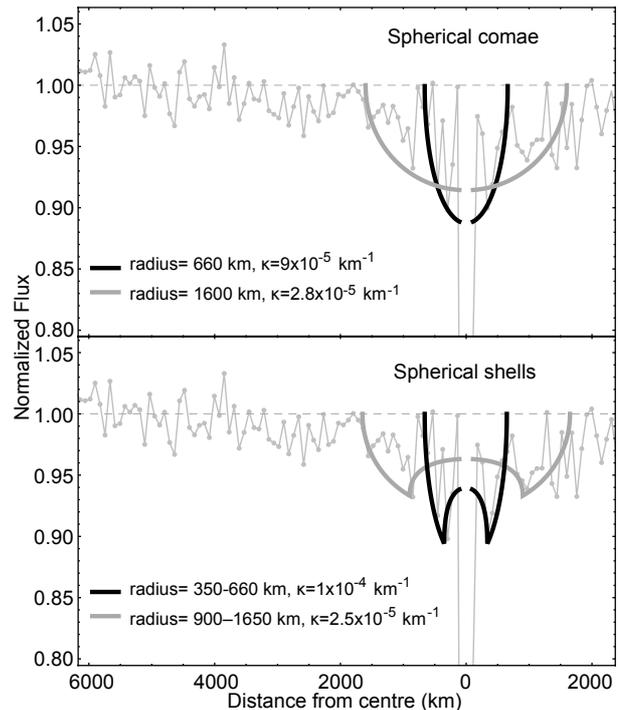

**Figure 9.** Example models for material around Chiron's nucleus, overlaying binned MORIS data. The data are binned in time over four points to better display the low-level, broad feature surrounding Chiron. The top plot shows models for spherical comae, filled homogeneously from the centre to the given radii. The bottom plot shows models for spherical shells, which are filled homogeneously in the listed radii ranges and are empty inside.

More clues to the centaur's structure are found in extended light-curve features. A few broad, shallow light-curve features were additionally identified in the 1993 and 1994 Chiron-occultation datasets. These features were thought to be due to extinction from a dust shell or other (possibly bound) coma material surrounding the nucleus. In 1993, the three broad features were aligned with each other as measured at different sites: they were 680, 2600, and 4700 km in extent and had line-of-sight optical depths 0.22–0.36 (B1, B2, and B4 in Bus *et al.* 1996). In 1994, the single broad feature was 750 km in width, centred on the midtime, and had line-of-sight optical depth 0.027 (F3 in Elliot *et al.* 1995). In 2011, there is also a low-level, midtime-centred, broad feature around the nucleus labelled B1 in Fig. 1, which becomes more apparent in binned light-curve data, at a significance level of 1.1 σ (see Fig. 9).

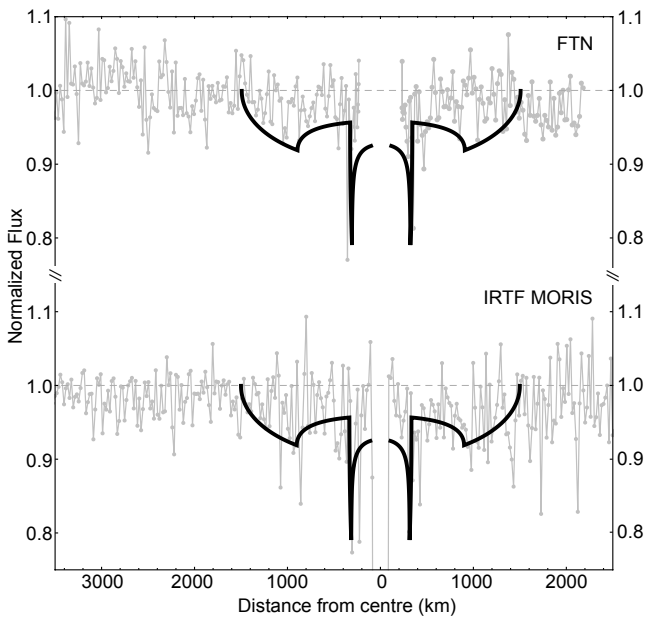

**Figure 10.** The MORIS and FTN data shown with a simple model of a thin shell of material at the location of the proposed ring system, from 314–334 km (Ortiz *et al.* 2015), plus a shell of material extending from 900–1500 km. The FTN data are binned by ten to be at the same resolution at the MORIS data.

To investigate different scenarios that could explain the broad light-curve feature in this dataset, a simple model of material surrounding the nucleus was created. The cases of a homogeneous sphere of material and a homogeneous spherical shell of material are considered, assuming a spherical nucleus on which the IRTF observations were a central chord. The relative change in stellar flux after a ray has travelled a distance $d$ into the material is $\exp(-\kappa d)$, where $\kappa$ is the extinction coefficient.

Figure 9 shows binned MORIS data along with example, model light-curve variations for spherical coma of two radii and for shells at two locations and thicknesses. These models generally match the observed extinction trend within a few thousand km of the nucleus; however, the signal rises more steeply than the models near the surface and there are more subtle variations in the observed flux than in the models. A reasonable match to both the FTN and MORIS data is shown in Fig. 10, in which a ring-like shell of $\kappa = 8 \times 10^{-4}$ km$^{-1}$ produces the sharp dips, and is combined with a wide shell from ~900–1500 km of $\kappa = 3.5 \times 10^{-5}$ km$^{-1}$ to produce the wings of the broad features. These models are representative only and are not fits to the data. Note that none of these scenarios reproduce the relative increase in transmission near the surface.

The results from this simple model suggest that (i) the sharp dips in the 2011 light curves (A1-A4, A8, and A10) are not caused by a full shell of material but have localized density (like planar rings, as proposed by Ortiz *et al.* 2015), because a shell would produce more extinction near the surface than that observed; (ii) there is a shell or diffuse ring material extending to approximately 1500 km from the centre (B1), with line-of-sight optical depth ~0.04 (based on the optical depth of FTN features: if this material were planar and not isotropic, the optical depth would be higher); and (iii) material inside of ~200 km has cleared out of the system.

Regarding (ii), the lack of spectral slope observed during the occultation (of Chiron alone) indicates that no significant micron-sized material was observed in the line-of-sight to the nucleus (in the reflected sunlight off of the nucleus or any residual flux from the star coming around the limb during the occultation). This dearth of material in front of the nucleus, and the shape of the light curves as shown in Fig. 9, is more consistent with a shell or diffuse ring than a spherical coma. The extent and location of the more distant, diffuse material do not match those of previous observations, which is consistent with the idea that such material is generated by sporadic jets or outgassing.

Regarding (iii), Sicardy *et al.* (2019) recently proposed that the ring dynamics for small bodies with non-axisymmetric nuclei would result in material being quickly cleared inside of the 1:2 spin-orbit resonance (particles orbit once for every two rotations of the nucleus). Ortiz *et al.* (2015) argued that Chiron is not spherical because of its double-peaked rotational light curve and proposed an ellipsoidal nucleus with major and minor axes of 143 and 190 km. Chiron's 1:2 resonance location is ~203 km from the centre; therefore, it is possible that the observed relative increase in transmission of light near the surface is the result of this clearing mechanism.

A thorough analysis was carried out on the SpeX data, specifically to investigate any flux versus wavelength trends in the near-infrared that could be diagnostic of the particle sizes of occulting material. Other than the occultation by the nucleus, there were no significant features in the SpeX data. The SNR was too low to detect any trends in spectral slope. This lack of detection is consistent with the MORIS light-curve features being 10-30 km in extent while each SpeX exposure spanned nearly 50 km at a ~50% duty cycle. Nonetheless, future multi-wavelength observations are recommended. Particle scattering models should be applicable to SpeX data for Chiron occultations by brighter stars and measured at higher spatial resolution.

## 5 CONCLUSIONS

A successful observation of a stellar occultation by the centaur Chiron was made from two sites in 2011, using three instruments, and spanning visible and near-infrared wavelengths. From those data, we determine the following:
(i) a lower limit on the diameter of Chiron's nucleus of 160.2±1.3 km, assuming a spherical shape, which is consistent with previous measurements;
(ii) a square-well model is valid for deriving object sizes and timing from low-cadence occultation data;
(iii) the two, previously-reported, azimuthally continuous features (proposed rings) around Chiron are located

between 280–360 km and have τ=0.6–1 (at 298.5–302 km and 308–310.5 km from the nucleus and $\tau_N$=0.5–0.9 for the proposed ring orientation): their widths vary in azimuth between ~2–3 and 3–6 km, and they are somewhat similar to Chariklo's inner ring;

(iv) the relatively low level of extinction observed within ~200 km of the nucleus indicates that Chiron's azimuthally continuous features are not a shell of material and are rather more like rings or ring arcs: furthermore, this could be indicative of clearing within the 1:2 resonance as a result of a non-axisymmetric nucleus;

(v) a possible additional arc and four azimuthally incomplete features are identified, between 225 and 1080 km: if real, these could be caused by transient jets or outgassing;

(vi) a possible shell or diffuse ring of material around Chiron is detected from ~900–1500 km, with τ~0.04 ($\tau_N$~0.03), of which future high-quality data could provide confirmation;

(vii) the SNR of the SpeX data is too low to detect and characterize any features, although future observations are encouraged in order to measure particle sizes of surrounding material; and

(viii) features similar to Chariklo's outer ring are below the detection threshold for this dataset and could exist in the Chiron system.

## ACKNOWLEDGEMENTS

This work was partially supported the South African National Research Foundation. ASB and AAS were visiting astronomers at the Infrared Telescope Facility, which is operated by the University of Hawaii under contract NNH14CK55B with the National Aeronautics and Space Administration. The work includes observations obtained with LCOGT network, which is owned and operated by Las Cumbres Observatory and the SARA Observatory, which is owned and operated by the Southeastern Association for Research in Astronomy (saraobservatory.org). MW acknowledges support from NSF grant AST-1615917. Thanks to S.J. Bus who was part of the original data acquisition and analysis team. AMZ and JDR are currently MIT Lincoln Laboratory employees. No Laboratory funding or resources were used to produce the result/findings reported in this publication.